\documentstyle[times,pramana,floats,amsmath,wrapfig,graphicx]{ias}
\begin{document}
%%%%%%%%%%%%%%%%%%%%%%%%%%%%%%%%%%%%%%%%%%%%%%%%%%%%%%%%%%%%%%%%%%%%%%%%%%%%%%
%%%%%%%%%%%%%%%%%%%%%%%%%%%%%%%%%%%%%%%%%%%%%%%%%%%%%%%%%%%%%%%%%%%%%%%%%%%%%%
\pagestyle{empty}
\begin{flushleft}
\Large{SAGA-HE-171-01, TMU-NT-01-02  \hfill March 14, 2001}  \\
\end{flushleft}

\vspace{1.5cm}
\begin{center}
\LARGE{{\bf Parametrization of}} \\
\vspace{0.2cm}

\LARGE{{\bf Nuclear Parton Distributions}} \\

\vspace{1.2cm}
\Large{\ \ M. Hirai, S. Kumano $^*$}         \\
 
\vspace{0.2cm}
{Department of Physics, Saga University}         \\
\vspace{0.1cm}
{Saga 840-8502, Japan} \\

\vspace{0.6cm}
{\ \ and M. Miyama $^\dagger$}         \\
 
\vspace{0.2cm}
{Department of Physics, Tokyo Metropolitan University} \\
\vspace{0.1cm}
{Tokyo, 192-0397, Japan}  \\

\vspace{1.4cm}
 
\Large{Invited talk given at} \\

\vspace{0.2cm}
{International Symposium on Nuclear Physics} \\
\vspace{0.4cm}

{Mumbai, India, Dec. 18 -- 22, 2000} \\
{(talk on Dec. 20, 2000)}  \\
\end{center}
\vspace{0.9cm}
\vfill
\noindent
{\rule{6.0cm}{0.1mm}} \\
\vspace{-0.3cm}
\normalsize

\noindent
{* Email: 98td25@edu.cc.saga-u.ac.jp, kumanos@cc.saga-u.ac.jp. } \\

\vspace{-0.4cm}
\noindent
{$\dagger$ Email: miyama@comp.metro-u.ac.jp.} \\

\vspace{-0.4cm}
\noindent
{\ \, Information on their research is available at
             http://www-hs.phys.saga-u.ac.jp.}  \\

\vspace{+0.3cm}
\hfill
{\large to be published in proceedings}

\vfill\eject
\setcounter{page}{1}
\pagestyle{plain}
%%%%%%%%%%%%%%%%%%%%%%%%%%%%%%%%%%%%%%%%%%%%%%%%%%%%%%%%%%%%%%%%%%%%%%%%%%%%%%
%%%%%%%%%%%%%%%%%%%%%%%%%%%%%%%%%%%%%%%%%%%%%%%%%%%%%%%%%%%%%%%%%%%%%%%%%%%%%%
\title{Parametrization of Nuclear Parton Distributions}

\author{M. Hirai, S. Kumano $^*$}
\address{Department of Physics, Saga University,
         Honjo-1, Saga 840-8502, Japan}
\author{\vspace{-0.8cm} \ \\
        M. Miyama $^\dagger$}
\address{Department of Physics, Tokyo Metropolitan University,
         Tokyo, 192-0397, Japan}
\abstract{
Optimum nuclear parton distributions are obtained by analyzing
available experimental data on electron and muon deep inelastic
scattering (DIS). The distributions are given at $Q^2$=1 GeV$^2$
with a number of parameters, which are determined by
a $\chi^2$ analysis of the data. Valence-quark distributions
are relatively well determined at medium $x$, but they are slightly
dependent on the assumed parametrization form particularly at small $x$.
Although antiquark distributions are shadowed at small $x$,
their behavior is not obvious at medium $x$ from the $F_2$ data. 
The gluon distributions could not be restricted well by the inclusive
DIS data; however, the analysis tends to support the gluon shadowing
at small $x$. We provide analytical expressions and computer subroutines
for calculating the nuclear parton distributions, so that other 
researchers could use them for applications to other high-energy
nuclear reactions.
}

\maketitle

\vspace{0.1cm}
%%%%%%%%%%%%%%%%%%%%%%%%%%%%%%%%%%%%%%%%%%%%%%%%%%%%%%%%%%%%%%%%%%%%%%%%%%%%%%
%%%%%%%%%%%%%%%%%%%%%%%%%%%%%%%%%%%%%%%%%%%%%%%%%%%%%%%%%%%%%%%%%%%%%%%%%%%%%%
\section{Introduction}\label{intro}
\vspace{-0.1cm}

Parton distributions in the nucleon are now known accurately
by using many experimental data on lepton and hadron reactions.
The distributions are expressed by parameters which are
determined by a $\chi^2$ analysis of the data. The determination
of the distributions is important not only for understanding 
internal structure of the nucleon but also for calculating
other reaction cross sections. If they are precisely known,
it becomes possible to find a signature for new physics
beyond the current framework by detecting a deviation
from theoretical predictions. In the recent years,
the parametrization studies have been extended to polarized parton
distributions \cite{aac}. The situation is not as good as the unpolarized
one in the sense that the details of polarized antiquark distributions
cannot be determined only by the $g_1$ measurements. The polarized
gluon distribution is also not well determined. We should wait
for hadron collider data.

In some sense, the situation of nuclear parton distributions is
similar to this polarized case. The antiquark and gluon distributions
are not well determined at this stage. In investigating 
high-energy nuclear reactions, the parton distributions in the ``nucleon" 
are often used instead of those in a nucleus.
Namely, a nucleus is often assumed as a simple collection of nucleons.
It is unsatisfactory in the sense, for example, that precise nuclear
distributions have to be known in order to find a quark-gluon signature
in heavy-ion collisions. 

It is recognized that nuclear parton distributions are modified from
the ones in the nucleon by the measurements of
nuclear $F_2$ structure functions \cite{f2sum}.
According to the data of the nucleus-deuteron ratio $F_2^A/F_2^D$,
the structure functions show shadowing at small $x$,
antishadowing around $x \sim 0.15$, and depletion at $x \sim 0.6$.
The ratio tends to increase at large $x>0.8$.
The modification has been also investigated theoretically, and
major features are now understood \cite{f2sum}. 
It is, however, not straightforward to find the details of the
modification in each parton distribution because all the distributions
contribute to $F_2$ in principle. 
Although the determination of the nuclear distributions is important
for practical applications, it is unfortunate that there was no
$\chi^2$ analysis. Of course, there were some trials to produce
the parton distributions from the nuclear data, for example, 
in a model-dependent way \cite{saga} and in a model-independent way
by Eskola, Kolhinen, and Ruuskanen \cite{helsinki}.
Here, we intend to pioneer the $\chi^2$ analysis of nuclear parton
distributions without relying on any theoretical models \cite{nuclp1}.
We also try to provide analytical expressions and computer subroutines,
so that other researchers could use them for their studies.
This talk is based on the analysis results in Ref. \cite{nuclp1}.
The nuclear parton distributions are provided at a fixed $Q^2$ with
a number of parameters, which are then determined by the $\chi^2$
analysis of experimental data.
The data are restricted to the inclusive electron and muon deep
inelastic data at this stage. We try to include some hadron collider
data in a later version. 

This paper consists of the following.
In Sec.\,2, assumed functional forms are explained,
and our $\chi^2$ analysis method is explained in Sec.\,\ref{analysis}
The analysis results are shown in Sec.\,\ref{results}
We explain obtained parton distributions so that other researchers could
use them in Sec.\,\ref{usage}
The summary is given in Sec.\,\ref{summary}

\vspace{0.1cm}
%%%%%%%%%%%%%%%%%%%%%%%%%%%%%%%%%%%%%%%%%%%%%%%%%%%%%%%%%%%%%%%%%%%%%%%%%%%%%%
%%%%%%%%%%%%%%%%%%%%%%%%%%%%%%%%%%%%%%%%%%%%%%%%%%%%%%%%%%%%%%%%%%%%%%%%%%%%%%
\section{Parametrization}\label{paramet}
\vspace{-0.1cm}

Because the parton distributions are well determined in the nucleon
and maximum nuclear effects are typically 20\% for a medium size nucleus,
it is a good idea to parametrize the modification instead of
the distributions themselves. There is, however, a disadvantage
in this approach although it could be a subtle problem.
If the scaling variable $x$ is defined by $x=Q^2/(2 m \nu)$ with
the nucleon mass $m$ and the energy transfer $\nu$ in the electron
or muon scattering, there are finite distributions even at $x>1$. 
If the nuclear distributions are taken as $w(x,A,Z) \, f(x)$,
where $f(x)$ is a parton distribution in the nucleon and 
$w(x,A,Z)$ is the nuclear modification, there is no way to obtain
the large $x$ distributions at $x>1$ due to $f(x \ge 1)=0$.
In this paper, this issue is neglected because of much more advantages.
In any case, even if the large $x$ part ($x>1$) is included
in the initial distributions, there is no reliable theoretical tool
to evolve them to larger $Q^2$ at this stage. 
In this way, the nuclear distributions are given at a fixed $Q^2$
($\equiv Q_0^2$) as
\begin{equation}
f_i^A (x, Q_0^2) = w_i(x,A,Z) \, f_i (x, Q_0^2),
\label{eqn:apart}
\end{equation}
where $f_i$ is the type-$i$ parton distribution in the nucleon,
and $w_i$ expresses the nuclear modification. Because the distributions
in the nucleon ($f_i$) are well known from other studies,
we try to parametrize $w_i$.
Alternatively, we could parametrize the nuclear distributions
($f_i^A$) themselves. However, it could lead to unphysical results
easily because a variety of nuclear data are not available in
comparison with the nucleon case. 
We call $w_i$ a weight function. In the following, we discuss
its functional form.

First, we discuss the $A$ dependence of the weight function $w_i$.
In this paper, we do not try to investigate the details
of the $A$ dependence. Because our analysis seems to be the first
$\chi^2$ trial, we assume the following simple $A$ dependence.
According to Ref. \cite{ds}, any nuclear cross sections could be
written in terms of volume and surface contributions: 
$\sigma_A = A \, \sigma_v + A^{2/3} \sigma_s$.
Then, the cross section per nucleon becomes
$\sigma_A /A = \sigma_v + A^{-1/3} \sigma_s$.
Of course, $\sigma_v$ and $\sigma_s$ have nuclear dependence;
however, the $1/A^{1/3}$ dependence could be considered as
the leading factor. We leave the issue of detailed $A$ dependence as
our future topic.

Second, because the nuclear distributions have finite distributions
even at $x=1$ and the distributions vanish in the nucleon,
the weight functions should have a property,
$w_i(x,A,Z) \rightarrow \infty$ as $x \rightarrow 1$.
In order to explain this $x$ region, we introduce a function
$1/(1-x)^{\beta_i}$. The rest of the $x$ dependence is assumed
as a polynomial form in our analysis. 

In this way, the following quadratic functional form is taken:
\begin{equation}
\text{``quadratic type":} \ \ \ 
w_i(x,A,Z) =1+\left( 1 - \frac{1}{A^{1/3}} \right) 
          \frac{a_i (A,Z) +b_i x+c_i x^2}{(1-x)^{\beta_i}} ,
\label{eqn:quad-wi}
\end{equation}
as the simplest one which could explain the measured $F_2^A$ data.
The parameter $a_i$ is considered to be dependent on the ratio $Z/A$,
and the reason is explained in the end of this section.
Because of the quadratic $x$ dependence, there are certain 
restrictions. For example, the valence-quark distributions
show antishadowing at small $x$ if the $F_2$ depletion at medium $x$ 
is explained mainly by the valence-quark modification. This is
because of a strong restriction due to the baryon-number
conservation. On the other hand, this simple parametrization
could be sufficient for the antiquark and gluon distributions,
where the detailed $x$ dependence at medium and large $x$
is not important at this stage by considering available data. 

Because the quadratic form could be too simple to explain
the data, we prepare the following second type:
\begin{equation}
\text{``cubic type":} \ \ \ 
w_i(x,A,Z)  = 1+\left( 1 - \frac{1}{A^{1/3}} \right) 
          \frac{a_i (A,Z) +b_i x+c_i x^2 +d_i x^3}{(1-x)^{\beta_i}} .
\end{equation}
There is an additional term $d_i x^3$. 
It is the advantage of this functional form that the distribution
shapes become more flexible to explain the experimental data.
For example, there is no aforementioned valence-quark problem at small $x$.
Namely, the valence distribution could show either shadowing or
antishadowing. However, the drawback is that it takes more
computing time in the $\chi^2$ analysis because of the additional freedoms.

As the distribution type $i$, we take valence up-quark,
valence down-quark, antiquark, and gluon distributions.
Flavor symmetric antiquark distributions are assumed here
due to the lack of data to discriminate the difference,
although there are some predictions on the asymmetry
in a nucleus \cite{skantiq}.
Furthermore, the valence up- and down-quark weight functions
are expected to be similar, so that the parameters are assumed
to be the same except for the constants $a_{u_v}$ and $a_{d_v}$.

There are three obvious conditions for the distributions:
nuclear charge, baryon number, and momentum.
In the parton model, they are expressed as 
\begin{alignat}{2}
& \text{charge} & \ \ \ \ 
Z & = \int dx \, A \left [   \frac{2}{3} (u^A - \bar u^A)
                    - \frac{1}{3} (d^A - \bar d^A) 
                    - \frac{1}{3} (s^A - \bar s^A) 
           \right ]
\nonumber \\
  & & &= \int dx \, \frac{A}{3} \left [ 2 \, u_v^A - d_v^A  \right ],
\label{eqn:charge} \\
& \text{baryon number} & \ \ \ \ 
A & = \int dx \, A \left [ \frac{1}{3} (u^A - \bar u^A)
                    + \frac{1}{3} (d^A - \bar d^A) 
                    + \frac{1}{3} (s^A - \bar s^A) 
           \right ]
\nonumber \\
  & & &= \int dx  \, \frac{A}{3} \left [  u_v^A + d_v^A  \right ],
\label{eqn:baryon} \\
& \text{momentum} & \ \ \ \ 
A & = \int dx \, A \, x  \left [ u^A + \bar u^A
                    + d^A + \bar d^A
                    + s^A + \bar s^A + g^A \right ]
\nonumber \\
  & & & = \int dx \, A \, x 
            \left [  u_v^A + d_v^A  + 6 \, \bar q^A + g^A \right ].
\label{eqn:momentum}
\end{alignat}
From these conditions, three parameters can be fixed. In our studies,
we decided to determine $a_{u_v}(A,Z)$, $a_{d_v}(A,Z)$, and $a_{g}(A,Z)$. 
Although $a_{\bar q}$ is still kept as a nuclear independent parameter,
these three parameters depend on a nucleus, in particular on
the ratio $Z/A$. Using the three conditions together with
the weight functions, we can express the parameters,
$a_{u_v}$, $a_{d_v}$, and $a_{g}$, in terms of 
nuclear independent constants and the ratio $Z/A$ \cite{nuclp1}.
Therefore, if the conditions are, for example, satisfied for the deuteron,
they are automatically satisfied for all the other isoscalar nuclei.
However, we also analyze nuclei with neutron excess.
Even if the three conditions are satisfied for the isoscalar nuclei, they
are not satisfied for other nuclei with different $Z/A$
factors. In this way, we introduce nuclear dependence in the parameters,
$a_{u_v}$, $a_{d_v}$, and $a_{g}$.

\vspace{0.1cm}
%%%%%%%%%%%%%%%%%%%%%%%%%%%%%%%%%%%%%%%%%%%%%%%%%%%%%%%%%%%%%%%%%%%%%%%%%%%%%%
%%%%%%%%%%%%%%%%%%%%%%%%%%%%%%%%%%%%%%%%%%%%%%%%%%%%%%%%%%%%%%%%%%%%%%%%%%%%%%
\section{$\bf \chi^2$ analysis method}\label{analysis}
\vspace{-0.1cm}

There are many experimental data in lepton and hadron reactions.
However, we use only the structure-function $F_2$ data in our
$\chi^2$ fit with the following reasons. First, 
because this is the first trial of the nuclear $\chi^2$ fit,
we would like to simplify the problem. Second, we would like to understand
how well the parton distributions are determined only by the inclusive
electron and muon scattering. Inclusion of other data such as Drell-Yan
measurements is left as our future research topic.

Nuclear $F_2$ measurements are usually shown by its ratio
to the deuteron $F_2$:
\begin{equation}
R^A_{F_2} (x,Q^2) \equiv \frac{F_2^A (x,Q^2)}{F_2^D (x,Q^2)}.
\end{equation}
In this paper, the leading order (LO) of $\alpha_s$ is used
in the analysis. The structure function $F_2$ is then expressed
in terms of the parton distributions as
\begin{align}
F_2^A (x,Q^2) & = \sum_q e_q^2 \, 
                  x \, [ q^A(x,Q^2) + \bar q^A(x,Q^2) ]
\nonumber \\
              & = \frac{x}{9} \, [ \, 4 \, u_v^A (x,Q^2) + d_v^A (x,Q^2) 
                               + 12 \, \bar q^A (x,Q^2) \, ],
\end{align}
where the number of flavor is assumed three. 
Theoretically, the above structure function and the deuteron $F_2$
can be calculated for a given set of parameters in the weight functions.
Then, the theoretical ratios $R^A_{F_2} (x,Q^2)$ are calculated and
they are compared with the corresponding experimental data
to obtain $\chi^2$:
\begin{equation}
\chi^2 = \sum_j \frac{(R_{F_2,j}^{A,data}-R_{F_2,j}^{A,theo})^2}
                     {(\sigma_j^{data})^2}.
\label{eqn:chi2}
\end{equation}
It should be noted that the parton distributions are provided
in the analytical form at a fixed $Q_0^2$, and the data are taken,
in general, at different $Q^2$ points. In order to calculate
the $\chi^2$, the distributions are evolved to the experimental
$Q^2$ points. This $Q^2$ evolution is calculated by 
the ordinary DGLAP (Dokshitzer-Gribov-Lipatov-Altarelli-Parisi) equations
in Ref. \cite{bf1}.

The initial $Q^2$ point is taken as $Q_0^2$=1 GeV$^2$ with the following
reason. In order to accommodate many data, a small $Q^2$ point is desirable.
However, it should be large enough to be within the perturbative QCD range.
As a compromise of these conditions, $Q_0^2$=1 GeV$^2$ is chosen.
Because we do not investigate the distributions in the nucleon,
it is necessary to employ a set of LO distributions.
There are three major groups in the
nucleon parametrization: CTEQ, GRV (Gl\"uck, Reya, and Vogt), 
and MRS (Martin, Roberts, and Stirling).
Among them, we decided to use the LO version of MRST-central gluon
in 1998 \cite{mrst} because analytical expressions are available
at $Q^2$=1 GeV$^2$ in the LO.

Before the actual analysis, we need to clarify the valence
up- and down-quark distributions in a nucleus in comparison
with those in the nucleon. If a nucleus consists of a simple
collection of nucleons, namely if there is no nuclear modification,
nuclear parton distributions are given by the summation of 
proton and neutron contributions:
$A f_i^A (x,Q^2)_{\text{no-mod}} = Z f_i^p (x,Q^2) + N f_i^n (x,Q^2)$.
Here, the nuclear parton distributions are defined
by the ones per nucleon. Isospin symmetry is assumed
in discussing the relation between
the distributions in the neutron and the ones in the proton.
Then, the deviation from this simple summation is expressed
by the weight functions:
\begin{align}
u_v^A (x,Q_0^2) & = w_{u_v} (x,A,Z) \, \frac{Z u_v (x,Q_0^2) 
                                         + N d_v (x,Q_0^2)}{A},
\nonumber \\
d_v^A (x,Q_0^2) & = w_{d_v} (x,A,Z) \, \frac{Z d_v (x,Q_0^2) 
                                         + N u_v (x,Q_0^2)}{A},
\nonumber \\
\bar q^A (x,Q_0^2) & = w_{\bar q} (x,A,Z) \, \bar q (x,Q_0^2),
\nonumber \\
g^A (x,Q_0^2)      & = w_{g} (x,A,Z) \, g (x,Q_0^2).
\label{eqn:wpart}
\end{align}

The technical part of the $\chi^2$ analysis is now ready, but we should also
specify used experimental data. As already mentioned, nuclear $F_2^A/F_2^D$
data are considered. Because of the DGLAP evolution equations,
the used data should be taken in a perturbative QCD region.
Therefore, the only data with $Q^2 \ge 1$ GeV$^2$ are used in the analysis.
We collected all the available data \cite{expdata}
by the European Muon Collaboration (EMC) at CERN,
the E49, E87, E139, and E140 Collaborations at SLAC,
the Bologna-CERN-Dubna-Munich-Saclay (BCDMS) Collaboration at CERN,
the New Muon Collaboration (NMC) at CERN,
and the E665 Collaboration at Fermilab.

%%%%%%%%%%%%%%%%%%%%%%%%%%%%%%%% figure %%%%%%%%%%%%%%%%%%%%%%%%%%%%%%%%%%%%%%
\begin{wrapfigure}{r}{0.46\textwidth}
\includegraphics[width=0.44\textwidth]{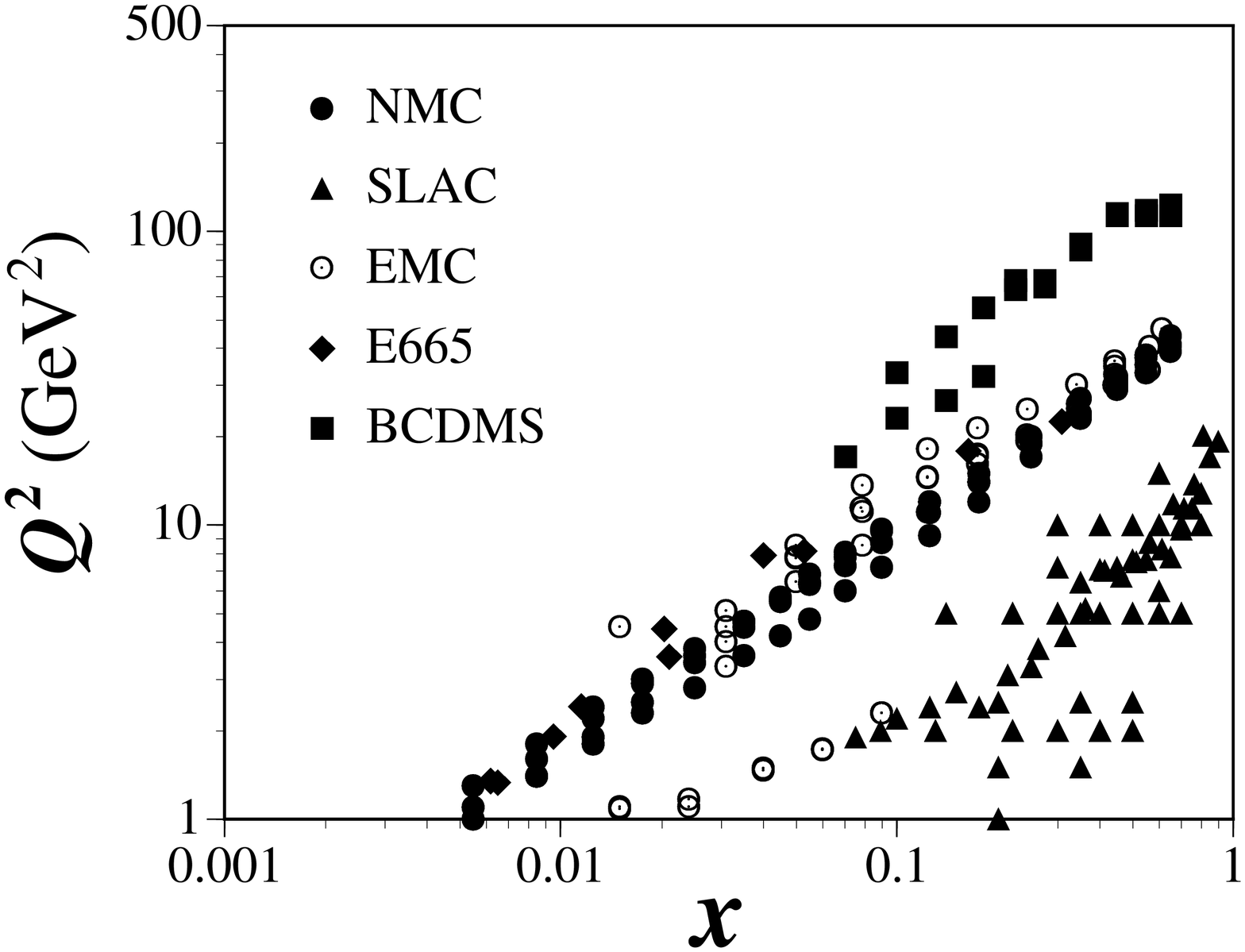}
\vspace{-0.8cm}
{\footnotesize Figure 1. $x$ and $Q^2$ values of the used data.}
\vspace{+0.9cm}
%%\vspace{-0.5cm}
%%\caption{\footnotesize Range of $x$ and $Q^2$ values of the used data.}
\label{fig:xq2}
\end{wrapfigure}
%%%%%%%%%%%%%%%%%%%%%%%%%%%%%%%% figure %%%%%%%%%%%%%%%%%%%%%%%%%%%%%%%%%%%%%%
Kinematical range of the experimental data is shown in Fig. 1.
The small $x$ data are taken by EMC, NMC, and E665, and they have rather
small $Q^2$ values in a restricted $Q^2$ range. This fact suggests that
it should be rather difficult to pin down the nuclear gluon distributions
from the scaling violation at small $x$. A significant number of data
are taken by the SLAC groups at large $x$ with relatively small $Q^2$
values. On the contrary, the BCDMS data are in the larger $Q^2$ range.
The data exist for the following nuclei:
helium (He), lithium (Li), beryllium (Be), carbon (C), nitrogen (N),
aluminum (Al),  calcium (Ca), iron (Fe), copper (Cu), silver (Ag), tin (Sn), 
xenon (Xe), gold (Au), and lead (Pb). The total number of the data is 309.
Theoretically, these nuclei are assumed as 
$^4$He, $^7$Li, $^9$Be, $^{12}$C, $^{14}$N, $^{27}$Al, $^{40}$Ca,
$^{56}$Fe, $^{63}$Cu, $^{107}$Ag, $^{118}$Sn,
$^{131}$Xe, $^{197}$Au, and $^{208}$Pb
in calculating the parton distributions.

\vspace{0.1cm}
%%%%%%%%%%%%%%%%%%%%%%%%%%%%%%%%%%%%%%%%%%%%%%%%%%%%%%%%%%%%%%%%%%%%%%%%%%%%%%
%%%%%%%%%%%%%%%%%%%%%%%%%%%%%%%%%%%%%%%%%%%%%%%%%%%%%%%%%%%%%%%%%%%%%%%%%%%%%%
\section{Results}\label{results}
\vspace{-0.1cm}

The $\chi^2$ analyses are done for both the quadratic and cubic
parametrization forms by the CERN subroutine {\sc Minuit} \cite{minuit}.
The following simplifications are introduced in the parameters. 
First, the parameter $\beta$ controls the large $x$ behavior.
Because the antiquark and gluon distributions do not contribute to
$F_2$ significantly at large $x$, the detailed values of 
$\beta_{\bar q}$ and $\beta_{g}$ are not important.
They are fixed at $\beta_{\bar q} = \beta_{g} =1$.
Furthermore, the relation between $b_g$ and $c_g$ is also fixed at
$b_g= -2 c_g$ because the gluon shape in the medium and large $x$
regions could not be determined reliably. Because of the same reason,
the gluon distribution is kept in the quadratic form even in the cubic
type analysis. The possible additional term $d_g x^3$ is associated with
the distribution shape at large $x$; however, such a term is almost
irrelevant in the present analysis.

The analysis results are shown for some of the used nuclei in
Figs. 2$-$7, where the dashed and solid curves
are the quadratic and cubic type results, respectively, at $Q^2$=5 GeV$^2$. 
We should be careful in comparing the theoretical curves with the data
because the data are taken in various $Q^2$ points, which are in general
different from $Q^2$=5 GeV$^2$. Considering even the $Q^2$ differences,
there are some deviations from the experimental data, for example,
in the medium-$x$ region of the carbon figure. However, if
we try to fit these data in this $x$ region, it is obvious that
the medium-$x$ depletion of the beryllium and gold is overestimated.
Therefore, this kind of small deviations seem to be inevitable in the
present $\chi^2$ analysis. 
Nevertheless, the figures indicate that our analyses are
reasonable in explaining the existing experimental data.
The obtained $\chi^2_{min}$ is 583.7 and 546.6 in the quadratic and
cubic analyses, respectively.

\vspace{-0.2cm}
%%%%%%%%%%%%%%%%%%%%%%%%%%%%%%%% figure %%%%%%%%%%%%%%%%%%%%%%%%%%%%%%%%%%%%%%
\noindent
\begin{figure}[h!]
\parbox[t]{0.46\textwidth}{
   \begin{center}
       \includegraphics[width=0.44\textwidth]{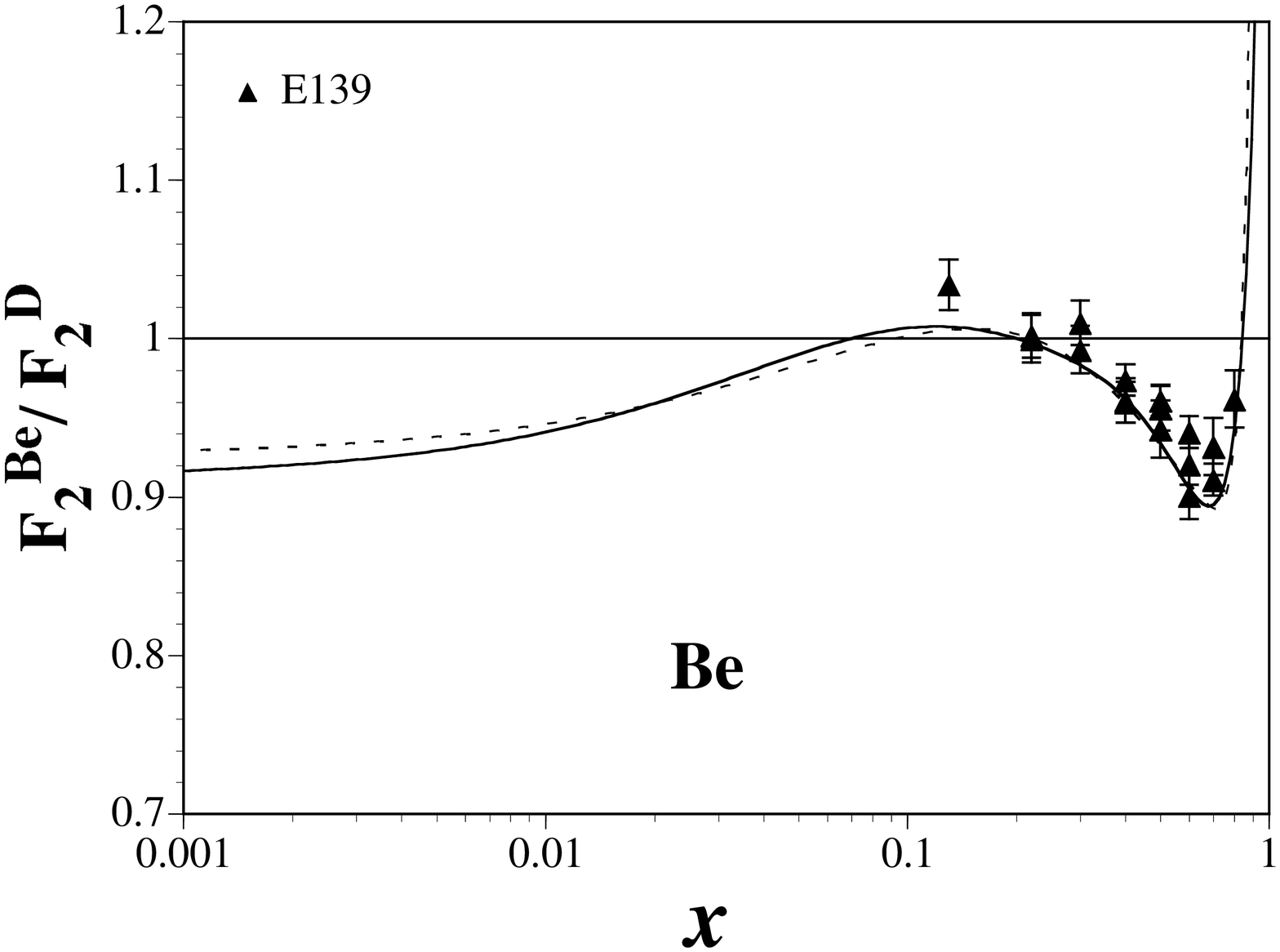}
   \end{center} 
\vspace{-0.6cm}
       {\footnotesize Figure 2. The dashed and solid curves are
                 fitting results for the beryllium
                 in the quadratic and cubic analyses, respectively,
                 at $Q^2$=5 GeV$^2$. They are compared with the data.}
       \label{fig:be}
}\hfill
\parbox[t]{0.46\textwidth}{
   \begin{center}
       \includegraphics[width=0.44\textwidth]{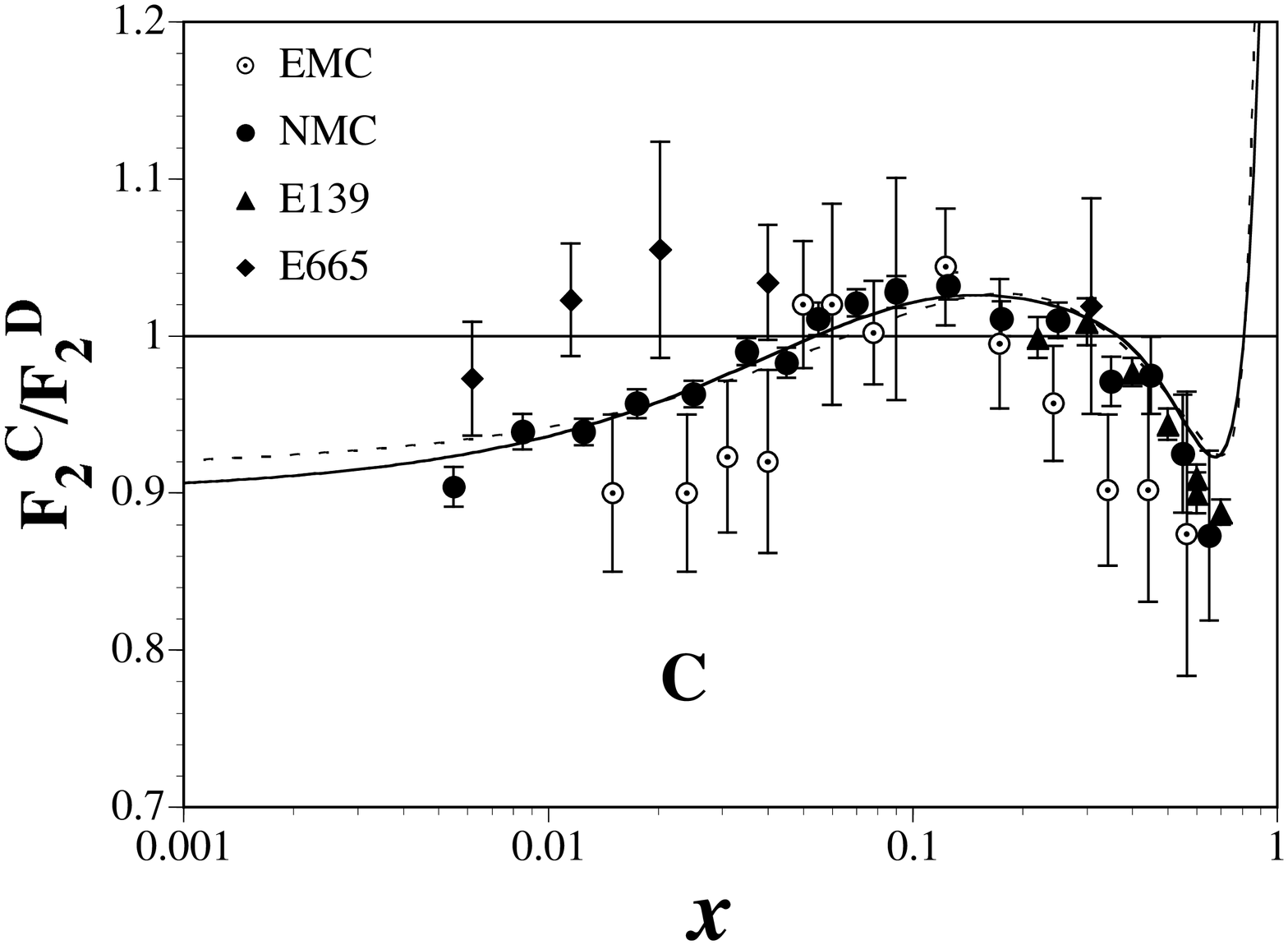}
   \end{center}
\vspace{-0.8cm}
   \begin{center}
       {\footnotesize Figure 3. Comparison with the carbon data.}
   \end{center}
       \label{fig:c}
}
%%%%%%%%%%%%%%%%%%%%%%%%%%%%%%%%%%%%%%%%%%%%%%%%%%%%%%
\vspace{-0.0cm}
\parbox[t]{0.46\textwidth}{
   \begin{center}
       \includegraphics[width=0.44\textwidth]{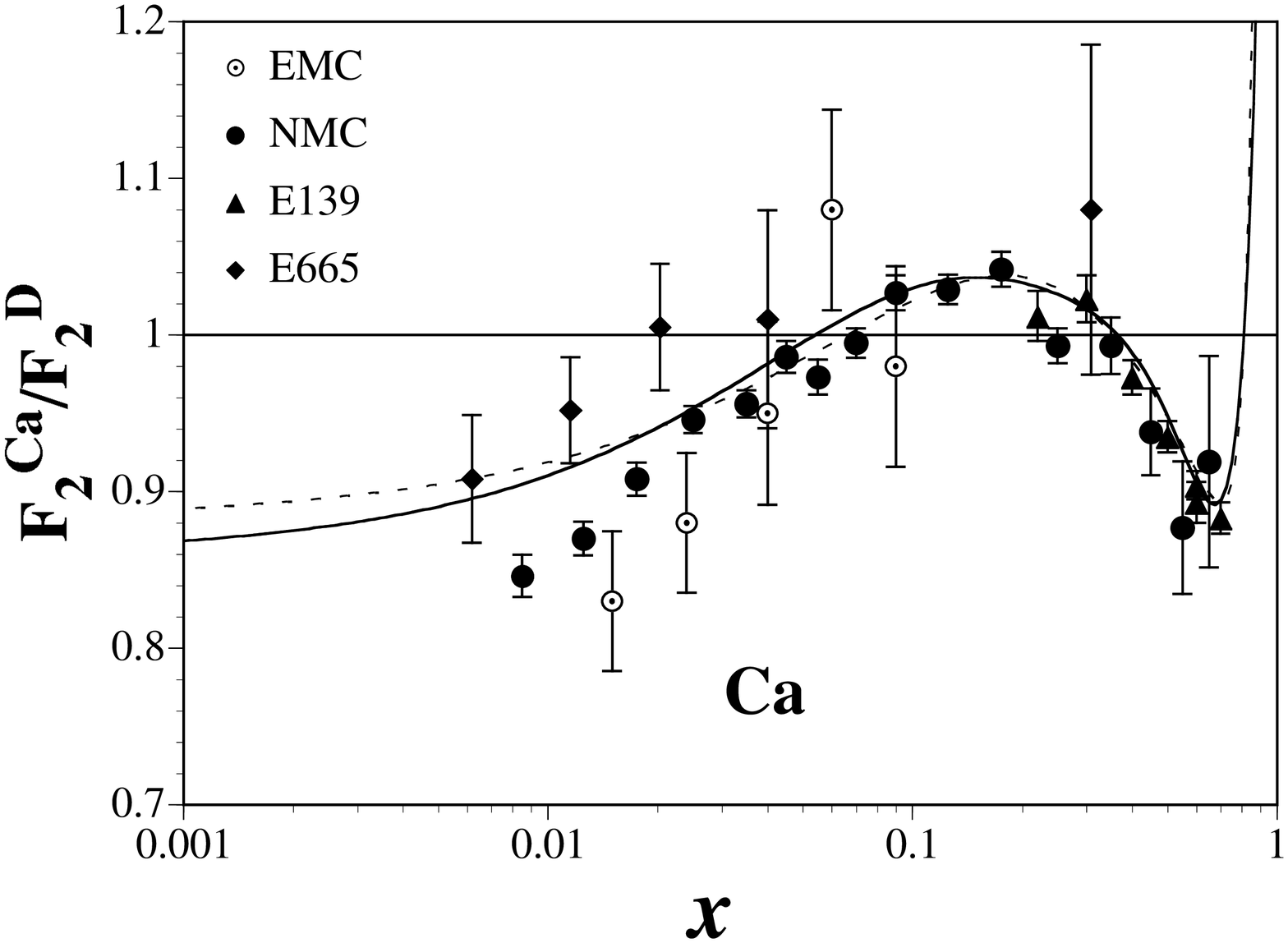}
   \end{center}
\vspace{-0.8cm}
   \begin{center}
       {\footnotesize Figure 4. Comparison with the calcium data.}
   \end{center}
       \label{fig:ca}
}\hfill
\parbox[t]{0.46\textwidth}{
   \begin{center}
       \includegraphics[width=0.44\textwidth]{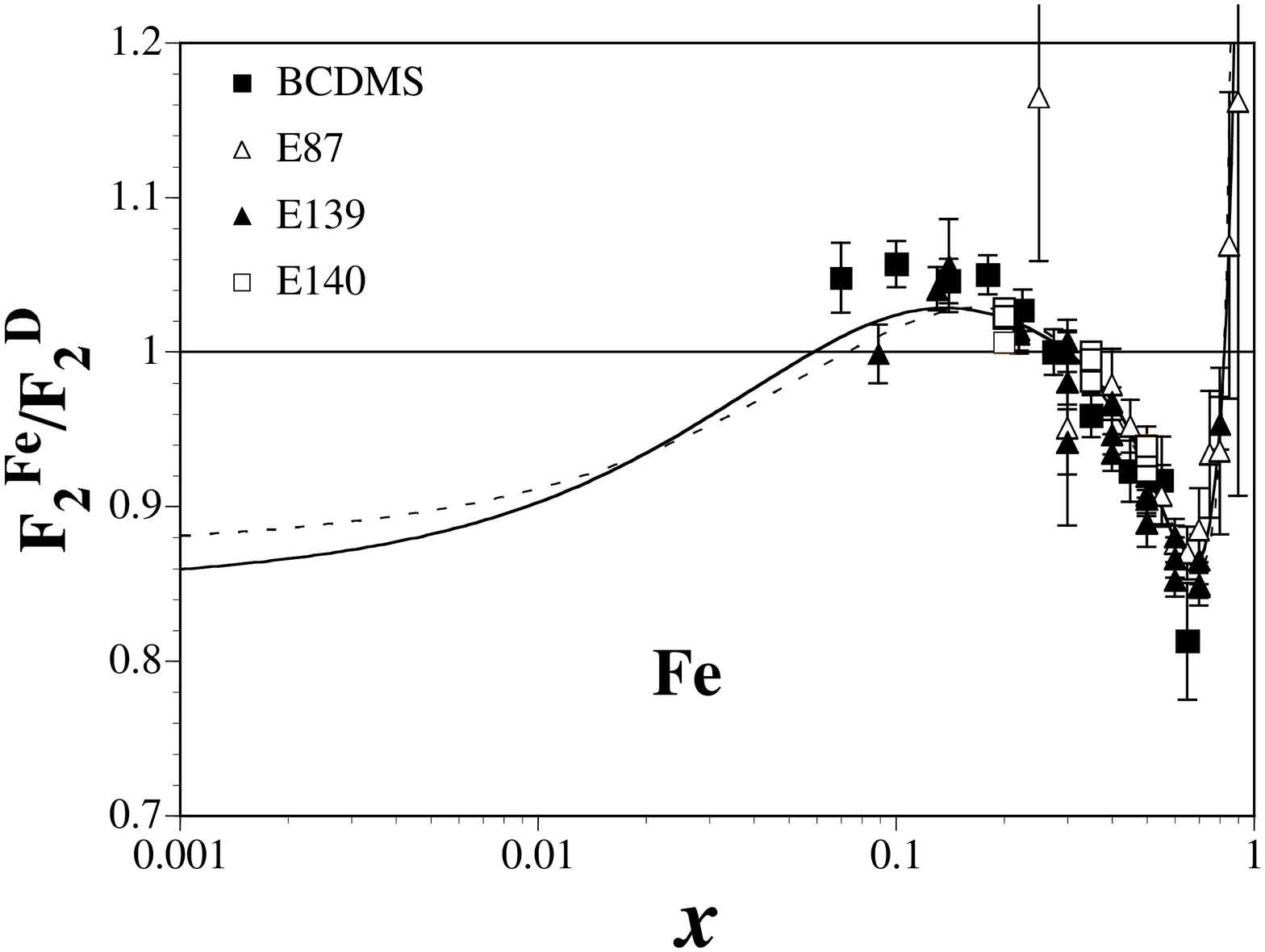}
   \end{center}
\vspace{-0.8cm}
   \begin{center}
       {\footnotesize Figure 5. Comparison with the iron data.}
   \end{center}
       \label{fig:fe}
}
%%%%%%%%%%%%%%%%%%%%%%%%%%%%%%%%%%%%%%%%%%%%%%%%%%%%%%
\vspace{+0.2cm}
\parbox[t]{0.46\textwidth}{
   \begin{center}
       \includegraphics[width=0.44\textwidth]{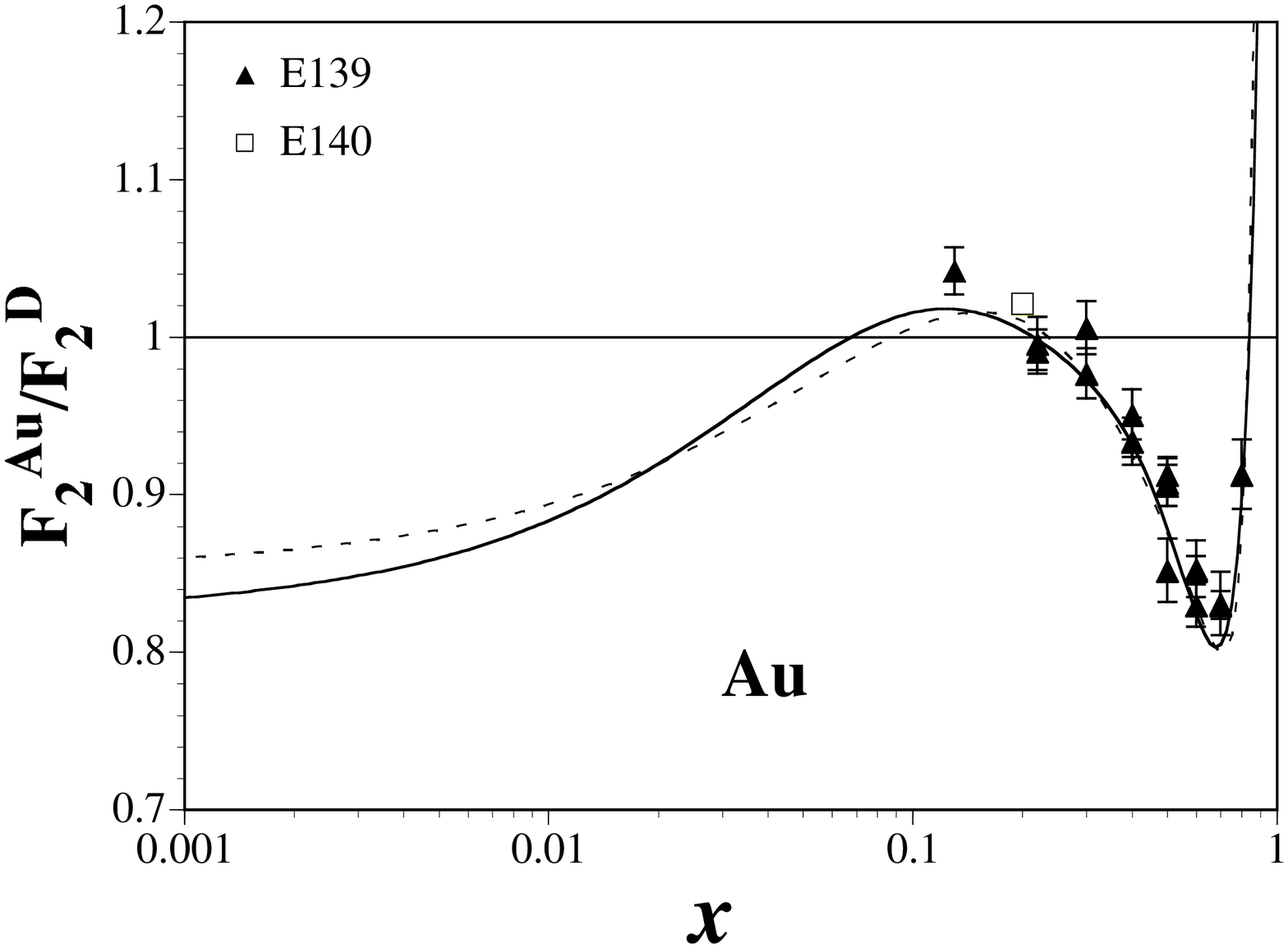}
   \end{center}
\vspace{-0.8cm}
   \begin{center}
   {\footnotesize Figure 6.Comparison with the gold data.}
   \end{center}
       \label{fig:au}
}\hfill
\parbox[t]{0.46\textwidth}{
   \begin{center}
       \includegraphics[width=0.44\textwidth]{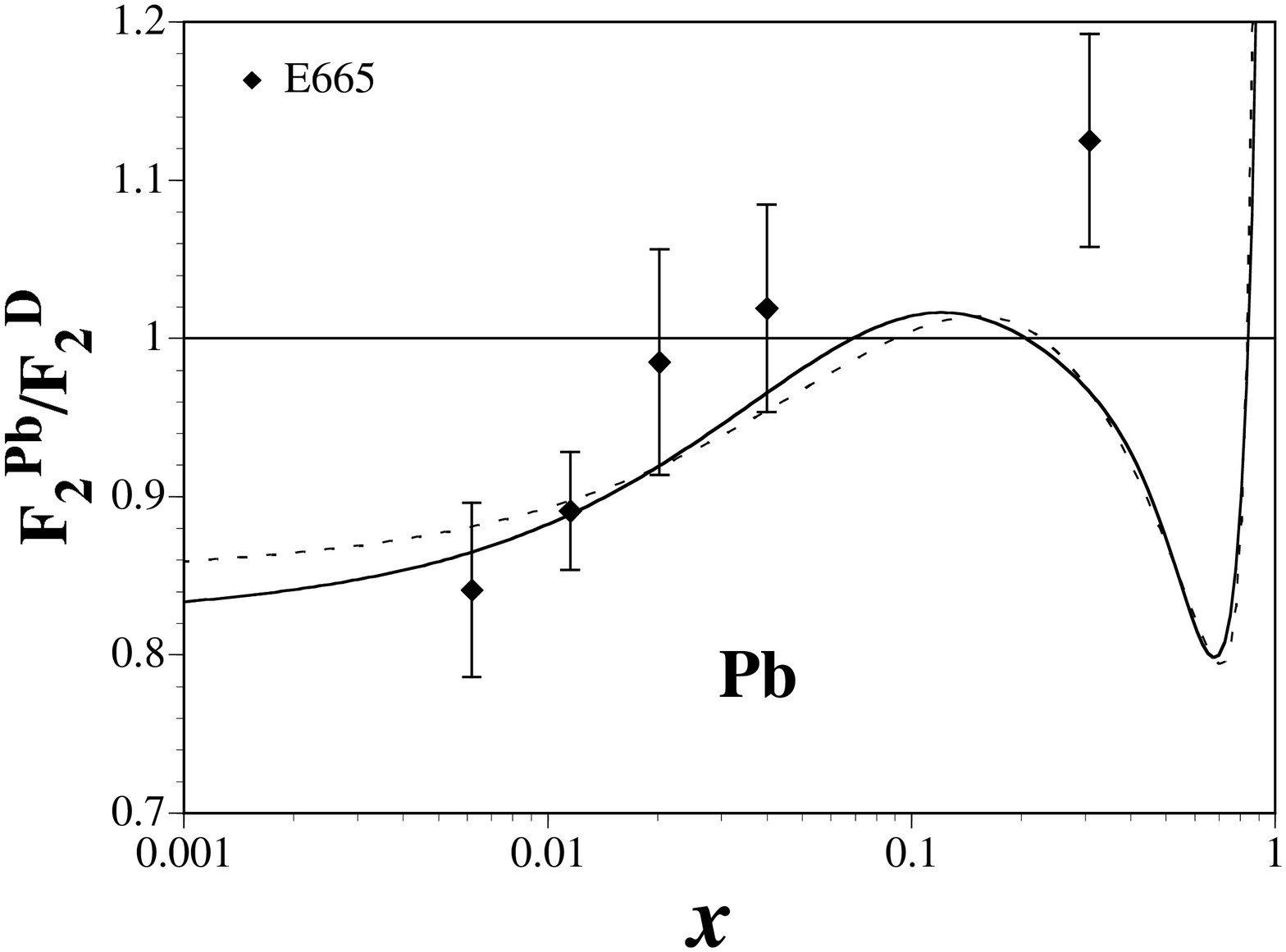}
   \end{center}
\vspace{-0.8cm}
   \begin{center}
      {\footnotesize Figure 7. Comparison with the lead data.}
   \end{center}
       \label{fig:pb}
}
\end{figure}
%%%%%%%%%%%%%%%%%%%%%%%%%%%%%%%% figure %%%%%%%%%%%%%%%%%%%%%%%%%%%%%%%%%%%%%%

There are systematic differences between the dashed and solid
curves in Figs. 2$-$7 at small $x$, where there are not many
experimental data. The quadratic results are in general above
the cubic ones at $x<0.01$, and they are below in the region,
$0.03 <x<0.14$. Because of the additional parameters, the
cubic analysis has more freedoms to readjust the distributions.
This fact results in such differences. In the medium and large
$x$ regions, both results are almost the same.
Of course, the cubic results are better than the quadratic ones
because of smaller $\chi^2_{min}$. However, as far as we see
in the figures, there are not so much differences between both
curves in comparison with the data,  so that both results could
be taken as possible nuclear parton distributions.

\vspace{-0.4cm}
%%%%%%%%%%%%%%%%%%%%%%%%%%%%%%%% figure %%%%%%%%%%%%%%%%%%%%%%%%%%%%%%%%%%%%%%
\noindent
\begin{figure}[h!]
\parbox[t]{0.46\textwidth}{
   \begin{center}
       \includegraphics[width=0.44\textwidth]{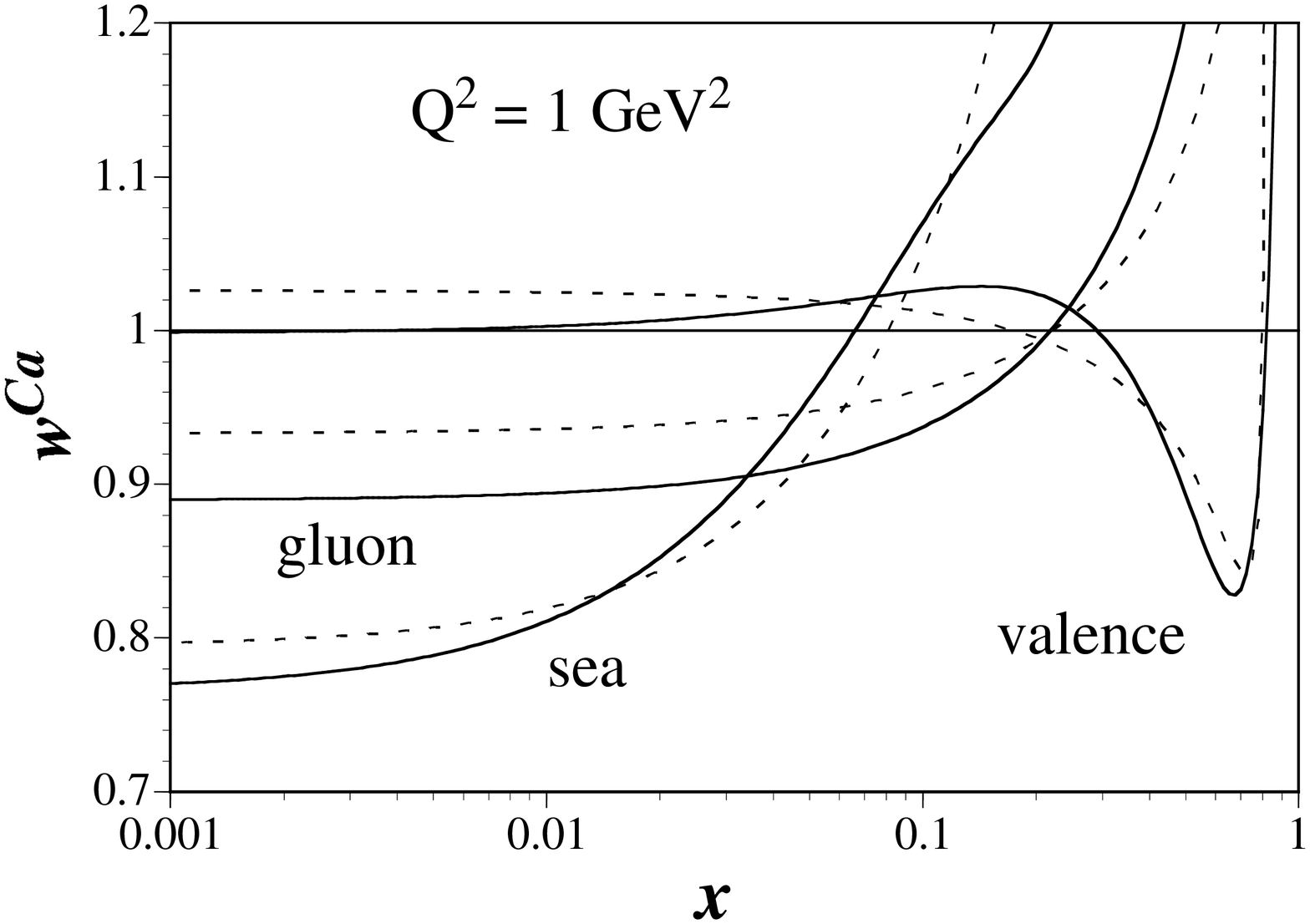}
       \vspace{-0.5cm}
    \end{center}
\vspace{-0.0cm}
{\footnotesize Figure 8. Weight functions are shown for the calcium nucleus.
                         The dashed and solid curves are the results for
                         the quadratic and cubic analyses, respectively. }
\vspace{+0.0cm}
\label{fig:wax}
}\hfill
\parbox[t]{0.46\textwidth}{
   \begin{center}
       \includegraphics[width=0.44\textwidth]{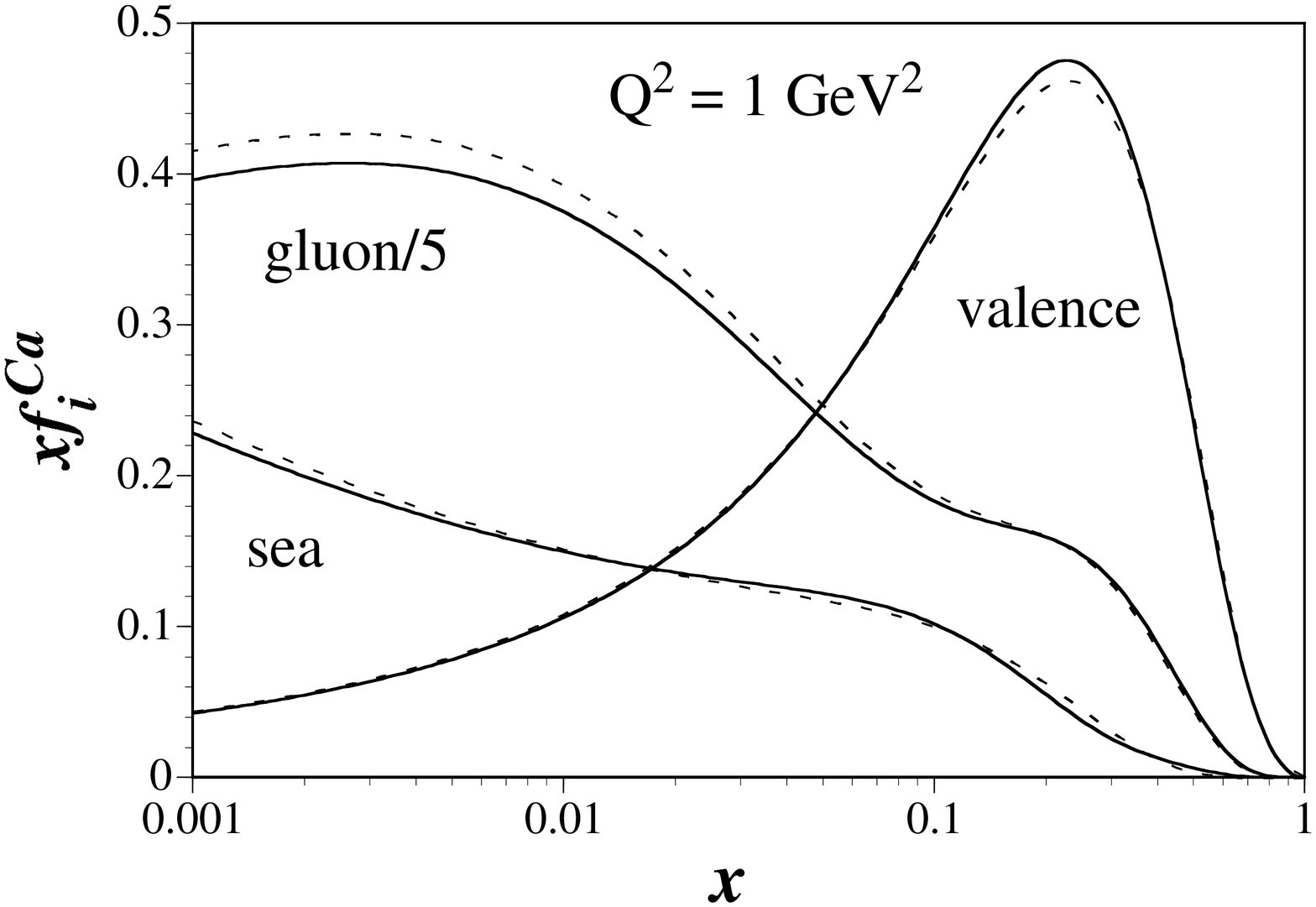}
       \vspace{-0.5cm}
   \end{center}
{\footnotesize Figure 9. Obtained parton distributions are shown
                         for the calcium nucleus.
                         The dashed and solid curves are the results for
                         the quadratic and cubic analyses, respectively. }
\vspace{+0.0cm}
\label{fig:fax}
}
\end{figure}
%%%%%%%%%%%%%%%%%%%%%%%%%%%%%%%% figure %%%%%%%%%%%%%%%%%%%%%%%%%%%%%%%%%%%%%%
\vspace{+0.2cm}

Using the obtained parameters, we plot the weight functions
for the calcium nucleus at $Q^2$=1 GeV$^2$ in Fig. 8.
The quadratic and cubic analysis results are shown by
the dashed and solid curves, respectively.
The valence-quark distributions are relatively well determined
from the $F_2$ data. However,
we notice that the small-$x$ behavior is slightly dependent on the
assumed functional form by comparing the dashed and solid curves.
In the quadratic fit, it shows antishadowing as
expected from the baryon-number conservation. However, it indicates slight
shadowing at very small $x$ ($\sim 0.001$) in the cubic fit.
This kind of issue cannot be solved only by the $F_2$ data,
and future neutrino factory data
for $F_3$ \cite{nuf3} should clarify the problem. 

The antiquark distributions are determined well at small $x$; however
their shapes are difficult to be determined at medium and large $x$. 
Even if the cubic functional form is taken for the antiquark distribution,
the obtained weight function is similar to the quadratic one according
to Fig. 8. The antiquark functions monotonically increase
as $x$ becomes larger.
The gluon functional shapes are similar to the antiquark functions.
The gluon distributions tend to be shadowed at small $x$; however, 
its determination is not easy in the larger $x$ region. 

Using the weight functions in Fig. 8 and the MRST distributions,
we obtain the parton distributions in the calcium nucleus in Fig. 9.
The dashed and solid curves indicate the distributions at $Q^2$=1 GeV$^2$
in the quadratic and cubic analyses, respectively. 
The quark distributions are well constrained by the $F_2$ measurements;
however, the gluon distributions are not reliably determined particularly
at large $x$. The details of the obtained distributions are 
explained in the next section, so that one could use them for one's
applications.

We should also mention the effects of our studies on the parton distributions
in the nucleon. Because the nuclear data are partially used in determining
the distributions in the nucleon without the nuclear corrections, the
existing nucleon parametrizations should be modified, particularly
in the valence-quark part.

%%\vspace{0.1cm}
%\vspace{-0.6cm}
%%%%%%%%%%%%%%%%%%%%%%%%%%%%%%%%%%%%%%%%%%%%%%%%%%%%%%%%%%%%%%%%%%%%%%%%%%%%%%
%%%%%%%%%%%%%%%%%%%%%%%%%%%%%%%%%%%%%%%%%%%%%%%%%%%%%%%%%%%%%%%%%%%%%%%%%%%%%%
\section{Parton distributions for practical usage}\label{usage}
\vspace{-0.1cm}

We provide both distributions obtained by the quadratic and
cubic analyses because the $\chi_{min}^2$ values are not much different.
However, the cubic type distributions are preferred because
the $\chi_{min}^2$ is smaller. We call the cubic and quadratic
distributions type \uppercase\expandafter{\romannumeral 1} and
type \uppercase\expandafter{\romannumeral 2}, respectively.
The obtained distributions are provided in two different ways:
analytical expressions and computer codes for numerical calculations.

\vspace{-0.2cm}
%%%%%%%%%%%%%%%%%%%%%%%%%%%%%%%%%%%%%%%%%%%%%%%%%%%%%%%%%%%%%%%%%%%%%%%%%%%%%%
\subsection{Analytical expressions}\label{analytical}
\vspace{-0.1cm}

First, analytical expressions are useful if one has own $Q^2$ evolution
program or if the $Q^2$ dependence could be neglected. 
The nuclear parton distributions are given by the weight functions and
MRST-LO (central gluon) distributions \cite{mrst}.
Here, we provide the expressions for the weight functions.
One should note that these functions are given at $Q^2$=1 GeV$^2$:
\begin{align}
\text{Type \uppercase\expandafter{\romannumeral 1}: cubic fit} &
\nonumber \\
w_{u_v}  & = 1+\left( 1 - \frac{1}{A^{1/3}} \right) \, 
          \frac{a_{u_v}(A,Z) +0.6222 x - 2.858 x^2 
                             +2.557 x^3}{(1-x)^{0.8107}} ,
\nonumber \\
w_{d_v}  & = 1+\left( 1 - \frac{1}{A^{1/3}} \right) \,
          \frac{a_{d_v}(A,Z) +0.6222 x - 2.858 x^2
                             +2.557 x^3}{(1-x)^{0.8107}} ,
\nonumber \\
w_{\bar q} & = 1+\left( 1 - \frac{1}{A^{1/3}} \right) \,
          \frac{-0.3313 +6.995 x -34.17 x^2 
                                 +62.54 x^3}{1-x} ,
\nonumber \\
w_{g}      & = 1+\left( 1 - \frac{1}{A^{1/3}} \right) 
          \frac{a_{g}(A,Z) +0.8008 x -0.4004 x^2}{1-x} ,
\end{align}
\begin{align}
\text{Type \uppercase\expandafter{\romannumeral 2}: quadratic fit} &
\nonumber \\
w_{u_v}     = & \, 1+\left( 1 - \frac{1}{A^{1/3}} \right) 
          \frac{a_{u_v}(A,Z) -0.2593 \, x + 0.2586 \, x^2}{(1-x)^{2.108}} ,
\nonumber \\
w_{d_v}     = & \, 1+\left( 1 - \frac{1}{A^{1/3}} \right) 
          \frac{a_{d_v}(A,Z) -0.2593 \, x + 0.2586 \, x^2}{(1-x)^{2.108}} ,
\nonumber \\
w_{\bar q}  = & \, 1+\left( 1 - \frac{1}{A^{1/3}} \right) 
          \frac{-0.2900 +3.774 \, x -2.236 \, x^2}{1-x} ,
\nonumber \\
w_{g}       = & \, 1+\left( 1 - \frac{1}{A^{1/3}} \right) 
          \frac{a_{g}(A,Z) + 0.4798 \, x -0.2399 \, x^2}{1-x} .
\end{align}
The actual values of $a_{u_v}$, $a_{d_v}$, and $a_{g}$ are not
provided here. One may determine them by one's effort so as to
satisfy the conditions in Eqs. (\ref{eqn:charge}), (\ref{eqn:baryon}),
and (\ref{eqn:momentum}). If one is considering a nucleus which
is one of the analyzed nuclei (D, He, Li, ..., Pb) in this paper,
one may simply take the tabulated values in Ref. \cite{nuclp1}.
For other nucleus, one is asked to follow the instructions
in Appendix of Ref. \cite{nuclp1}. However, the requested
nucleus should not be too far away from the analyzed nuclei.
Because the distributions are given at $Q^2$=1 GeV$^2$, one should
evolve them to the appropriate $Q^2$ point in one's project.

\vspace{-0.3cm}
%%%%%%%%%%%%%%%%%%%%%%%%%%%%%%%%%%%%%%%%%%%%%%%%%%%%%%%%%%%%%%%%%%%%%%%%%%%%%%
\subsection{Computer subroutines}\label{library}
\vspace{-0.1cm}

If one thinks that it is tedious to evolve the analytical expressions,
one had better use computer subroutines, which are prepared to 
calculate the nuclear parton distribution at any given $x$ and $Q^2$
points. The variables $x$ and $Q^2$ are divided into small steps,
and a grid data is prepared for each nucleus at these $x$ and $Q^2$ points.
The linear interpolation is used for $log \, Q^2$ because the $Q^2$
dependence is small, and the cubic Spline interpolation is used
for the $x$ part. Suggested kinematical ranges are $10^{-9} \le x \le 1$
and $1 \ {\rm GeV}^2 \le Q^2 \le 10^5 \ {\rm GeV}^2$. Our codes could
be used for calculating the distributions in other nuclei than
the analyzed ones. The detailed instructions are found in the web page
of Ref. \cite{nucl-lib}.

\vspace{0.1cm}
%%%%%%%%%%%%%%%%%%%%%%%%%%%%%%%%%%%%%%%%%%%%%%%%%%%%%%%%%%%%%%%%%%%%%%%%%%%%%%
%%%%%%%%%%%%%%%%%%%%%%%%%%%%%%%%%%%%%%%%%%%%%%%%%%%%%%%%%%%%%%%%%%%%%%%%%%%%%%
\section{Summary}\label{summary}
\vspace{-0.1cm}

We have done $\chi^2$ analyses of nuclear structure-function ratios
$F_2^A/F_2^D$ by collecting existing electron and muon deep inelastic
experimental data. Assuming simple $1/A^{1/3}$ dependence in the 
nuclear modification part, we parametrized the initial nuclear parton
distributions at $Q^2$=1 GeV$^2$. They are taken as the quadratic
or cubic functional form with a number of parameters, which are
then determined by the $\chi^2$ analysis. We have obtained reasonable
fit to the data. As a result, the valence-quark distributions are
reasonably well determined except for the small $x$ region. 
The obtained antiquark distributions indicate shadowing at small $x$.
However, the antiquark and gluon distributions are not well fixed by the
$F_2$ data in the medium and large $x$ regions. In particular,
it is difficult to determine the gluon distributions in the whole $x$ region.
However, the results indicate the gluon shadowing at small $x$.

\vspace{0.1cm}
%%%%%%%%%%%%%%%%%%%%%%%%%%%%%%%%%%%%%%%%%%%%%%%%%%%%%%%%%%%%%%%%%%%%%%%%%%%%%%
%%%%%%%%%%%%%%%%%%%%%%%%%%%%%%%%%%%%%%%%%%%%%%%%%%%%%%%%%%%%%%%%%%%%%%%%%%%%%%
\section*{Acknowledgments}
\vspace{-0.1cm}

The authors were supported by the Grant-in-Aid for Scientific Research
from the Japanese Ministry of Education, Culture, Sports, Science,
and Technology. M.H. and M.M. were supported by the JSPS Research Fellowships
for Young Scientists. 
S.K. would like to thank the chairperson,
B. K. Jain, and other organizers of this conference for their invitation
and for taking care of his stay in Mumbai.
He also thanks the Institute for Nuclear Theory in Seattle
for its hospitality and the US Department of Energy for partial support
in writing up this paper.

%%%%%%%%%%%%%%%%%%%%%%%%%%%%%%%%%%%%%%%%%%%%%%%%%%%%%%%%%%%%%%%%%%%%%%%%%%%%%%
\vspace{0.6cm}
\noindent
{* Email: 98td25@edu.cc.saga-u.ac.jp, kumanos@cc.saga-u.ac.jp;} \\

\vspace{-0.4cm}
\noindent
{\ \, WWW: http://www-hs.phys.saga-u.ac.jp.}  \\

\vspace{-0.4cm}
\noindent
{$\dagger$ Email: miyama@comp.metro-u.ac.jp.} \\

%%%%%%%%%%%%%%%%%%%%%%%%%%%%%%%%%%%%%%%%%%%%%%%%%%%%%%%%%%%%%%%%%%%%%%%%%%%%%%%
%%%%%%%%%%%%%%%%%%%%%%%%%%%%%%%%%%%%%%%%%%%%%%%%%%%%%%%%%%%%%%%%%%%%%%%%%%%%%%%

%%%%%%%%%%%%%%%%%%%%%%%%%%%%%%%%%%%%%%%%%%%%%%%%%%%%%%%%%%%%%%%%%%%%%%%%%%%%%%%
%%%%%%%%%%%%%%%%%%%%%%%%%%%%%%%%%%%%%%%%%%%%%%%%%%%%%%%%%%%%%%%%%%%%%%%%%%%%%%%


\begin{thebibliography}{99}
\bibitem{expdata} Because it is too lengthy to list all the experimental
                  papers, the authors suggest that the reader look at
                  the reference section of Ref. \cite{nuclp1}.
\end{thebibliography}
\end{document}